\documentclass[%
preprint,
 bibnotes,
 amsmath,amssymb,
 11pt
]{revtex4-2}

\usepackage{graphicx}
\usepackage{dcolumn}
\usepackage{bm}
\usepackage{xcolor}
\usepackage{color,soul}
\usepackage[11pt]{moresize}
\usepackage{inputenc}

\begin{document}


\title{Crossing of the branch cut: the topological origin of a universal $2\pi$-phase retardation in non-Hermitian metasurfaces}


\author{R\'{e}mi Colom$^1$, 
Elena Mikheeva$^{2}$,
Karim Achouri$^{3}$,
Jesus Zuniga-Perez$^{2,4}$,
Nicolas Bonod$^{5}$,
Olivier J.F. Martin$^{3}$,
Sven Burger$^{1,6}$,
Patrice Genevet$^{2}$}

\email[Correspondence email address: patrice.genevet@crhea.cnrs.fr]{}
\affiliation{$^1$Zuse Institute Berlin, Takustraße 7, 14195 Berlin, Germany}
\affiliation{$^2$CNRS, CRHEA, Université Côte d’Azur, 06560 Valbonne, France}
\affiliation{$^3$Ecole Polytechnique Federale
de Lausanne, Lausanne, VD, Switzerland}
\affiliation{$^4$MajuLab, International Research Laboratory IRL 3654, CNRS, Université Côte d’Azur, Sorbonne Université,
National University of Singapore, Nanyang Technological University, Singapore, Singapore}
\affiliation{$^5$Aix-Marseille Univ, CNRS, Centrale Marseille, Institut Fresnel, 13397 Marseille, France}
\affiliation{$^6$JCMwave GmbH, Bolivarallee 22, 14050 Berlin, Germany}


\begin{abstract}

Full wavefront control by photonic components requires that the spatial phase modulation on an incoming optical beam ranges from $0$ to $2\pi$. Because of their radiative coupling to the environment, all optical components are intrinsically non-Hermitian systems, often described by reflection and transmission matrices with complex eigenfrequencies. Here, we show that Parity-Time symmetry breaking -either explicit or spontaneous- moves the position of Zero singularities of the reflection or transmission matrices from the real axis to the upper part of the complex frequency plane. A universal $0$ to $2\pi$-phase gradient of an output channel as a function of the real frequency excitation is thus realized whenever the discontinuity branch bridging a Zero and a Pole, $i.e.$ a pair of singularities, is crossing the real axis. This basic understanding is applied to engineer electromagnetic fields at interfaces, including, but not limited to, metasurfaces. Non-Hermitian topological features associated with exceptional degeneracies or branch cut crossing are shown to play a surprisingly pivotal role in the design of resonant photonic systems.


\end{abstract}

\maketitle


\section{\label{sec:level1}Introduction}

Over the last decades, the development of artificial materials and artificial interfaces to address arbitrarily output-from-inputs signals has considerably modernized the fields of optics and optical design. A particularly impressive amount of research in the field of metamaterials has been devoted to the design and realization of ultra-thin artificial optical surfaces for wavefront engineering and control. These optical surfaces, also dubbed metasurfaces, rely on the coherent scattering of light by a sizable distribution of nanoscatterers of various shapes and material compositions. The list of optical effects achieved using metasurfaces is extensive, ranging from anomalous reflection and refraction \cite{yu_light_2011,lalanne_blazed_1998, fattal_flat_2010, genevet_ultra-thin_2012, aieta_out_plane_2012,pors_broadband_2013,karimi_generating_2014, yang_dielectric_2014}, all the way to the design of utterly complex vectorial holographic surfaces \cite{wang_broadband_2018, chen_broadband_2018, ren_metasurface_2019, song_ptychography_2020, sawant_aberration-corrected_2021,song_broadband_2021,song_bandwidth-unlimited_2021}. The interested reader could refer to reviews and references therein for detailed descriptions and applications \cite{chen_review_2016, genevet_recent_2017, kamali_review_2018, chen_flat_2020}.
In practice, light control is achieved by varying the geometries of adjacent elements so as to provide spatially-varying phase retardation on the incoming wavefront. Surprisingly, among all physical mechanisms of interest for the design of these phase building blocks, including the Pancharatnam-Berry phase in anisotropic nanoparticles \cite{berry_adiabatic_1987, biener_formation_2002} and effective index waveguide modes in nanopillars \cite{lalanne_blazed_1998}, the fundamental origin of the $2\pi$-phase modulation associated with resonant scattering of Mie nanoparticles has not been deeply investigated. Ultimately, and despite all the efforts in understanding which physical mechanisms are leading to optimal designs, the most advanced patterns often require witless numerical parameter searches.\\

Here, we discovered that full $2\pi$-phase retardation commonly used for designing metasurfaces exclusively relies on the relative spectral positions of topological singularities of the metasurface response functions. These manifest as Zero-Pole pairs in the complex frequency plane. Resonant-phase metasurfaces are designed to connect N input $\mathbf{E_{in}}$-modes with N output $\mathbf{E_{out}}$-modes. They behave as regular linear open wave systems mathematically described by $N\times N$ response matrices $\mathbf{F}\left( \omega \right)$ such that, $\mathbf{E_{out}}=\mathbf{F}\left( \omega \right)\cdot\mathbf{E_{in}}$ \cite{sweeney_theory_2020,kang2021}, where $\mathbf{F}(\omega)$ represents either the S-matrix, the reflection matrix $\mathbf{R}$ or the transmission matrix $\mathbf{T}$ \footnotemark[1] \footnotetext[1]{While the results presented here can be extended to multiport cases, we consider for simplicity and throughout the rest of this paper only one input and one output modes. Regarding symmetry considerations discussed further, we fix the input incident angle at normal incidence and consider only the case of polarization preserving metasurfaces}. Zeros and Poles of response matrices or functions behave as phase singularities around which the phase is spiraling in a vortex-like way \cite{grigoriev_optimization_2013,grigoriev_singular_2013}. Metasurfaces, like most optical devices, are intrinsically non-Hermitian devices, suffering from scattering losses and potentially from intrinsic loss or gain. Therefore, their phase singularities generally occur in the complex frequency plane.  Singularities have long been known to play an important role in optics \cite{gbur2016singular,nye1999natural,soskin2001singular,dennis2009singular}. Here, we show how phase singularities occurring in the complex plane greatly influence the optical response of resonant metasurfaces. More precisely, we show that $2\pi$ phase accumulation needed for wavefront engineering of reflected or transmitted output channel has a deep topological origin. It requires two complementary singularities, known as Poles and Zeros, to be disposed in the complex frequency plane on either side of the real axis. This way, the branch-cut discontinuity bridging these two singularities crosses the real axis, providing $2\pi$ phase accumulation as a function of the real frequency excitation. We demonstrate that one can rely on symmetry-breaking in order to have two singularities of the same pair located on each side of the real axis which could induce $2\pi$ circulation for any path in the complex plane encircling a singularity. We propose two approaches inducing Parity-time symmetry breaking to move the Zeros from the real axis to the upper part of the complex plane.\

The first solution is relatively straightforward and relies on explicit symmetry-breaking arguments. The second solution, which is linked to the regime known as Huygens metasurfaces \cite{decker_highefficiency_2015}, is more subtle as it occurs in symmetric systems featuring spontaneously symmetry-broken states. We thus show that the physical origin of Huygens metasurfaces is deeply rooted in topological concepts. Finally, looking at the analytical expressions of metasurface boundary conditions, namely the "generalized sheet transition conditions" (GSTC), we identified the symmetry conditions of the electromagnetic modes to be considered for spontaneous symmetry-breaking. 

\section{\label{sec:level3} Complex frequency analysis and phase integration}
We begin our analysis by studying the analytical properties of a non-Hermitian metasurface represented by a response function $F(\omega)$, $e.g.$ its reflection $R$ or its transmission $T$ coefficients. Expanding $F(\omega)$ on the basis of its Poles $\omega_{\text{p},n}$ and Zeros $\omega_{\text{z},n}$ using Weierstrass  product expansion, we have \cite{grigoriev_optimization_2013, grigoriev_singular_2013}: 
\begin{equation}
F(\omega) = A\exp\left(iB\omega\right)\prod_{n = -\infty}^{\infty}\frac{\omega-\omega_{\text{z},n}}{\omega-\omega_{\text{p},n}},
\label{Weier_T}
\end{equation}
where A and B are constants depending on the properties of the metasurface and the considered response function. By definition, the phase is given by $ \text{Arg}(F) = \text{Im}\left(\log(F)\right)$ and, as a consequence, Zeros and Poles of $F$ correspond to logarithmic branch points at which the phase becomes singular \cite{arfken1999}. The phase is thus spiraling as vortices around Zeros and Poles of F in the same way as the phase of a complex number in  $\mathbb{C}$  rotates around the origin of the complex plane \cite{grigoriev_optimization_2013}. Zeros and Poles are thus topological defects \cite{mermin1979}. While the Poles of $\mathbf{S}$, $R$, and $T$ are identical and are associated with resonances or quasi-normal modes solutions of the wave equation, their complex Zeros are different and provide insights into the physical effects governing the metasurface response. In particular, they greatly influence the real-frequency response of metasurfaces. Zeros of the $\mathbf{S}$-matrix are complex conjugate of Poles  for passive lossless metasurfaces and can reach the real axis in lossy structures where they are linked to perfect coherent absorption  \footnotemark[2] \footnotetext[2]{Poles of $\mathbf{S}$-matrix are complex solutions of wave equation calculated with only outgoing boundary conditions, while Zeros of $\mathbb{S}$ are calculated with only in-going boundary conditions. Poles and Zeros of $S$-matrix are not independent: in the lossless system, they are always complex conjugates due to the time-reversal symmetry \cite{chong2010}. It is also not possible to bring Poles and Zeros to the real frequency axis without gain-loss engineering due to energy conservation. \cite{grigoriev_singular_2013}}. Zeros of $R$ and $T$ coefficients are linked to Reflectionless and Transmissionless states, respectively \footnotemark[3] \footnotetext[3]{Reflectionless Scattering Modes (RSM) \cite{sweeney_theory_2020} and (TSM) Transmissionless Scattering Modes \cite{kang2021} are often used in the literature to denote the two specific cases when Zeros are brought to the real axis, which could occur even in lossless systems. The rest of the time, i.e. for any value of the Zero-frequency in the complex plane, the states are called Reflectionless and respectively Transmissionless states}. Let us consider a toy-model  $F(\omega)=(\omega-\omega_z)/(\omega-\omega_p)$, derived from Eq. \eqref{Weier_T} for $A = 1$ and $B = 0$ and considering a single Zero-Pole pair, featuring only one Zero $\omega_z$ and one Pole $\omega_p$ located on each side of the real axis, as indicated by the blue and red regions respectively in Fig.~\ref{fig:Gauss_Symm}(a.). We define as topological charge the quantity $q$ calculated by evaluating the winding number along a counterclockwise contour encircling the singularities as\cite{mermin1979, shen_topological_2018}: 
\begin{equation}
\label{eq:Qcharge}
q = \frac{1}{2\pi}\oint_{C_l}\frac{d\text{Arg}\left(F(\omega)\right)}{d\omega}  d\omega,
\end{equation}
equalling to $q = +1$ for a Zero and $q = -1$ for a Pole (see SI). The phase varies by $2\pi$ around the Zeros and the Poles with an opposite sign. As a result, the phase variation around both is null, see Fig.~\ref{fig:Gauss_Symm}(b.).  Moreover, Zeros and Poles occur in pairs and are connected by a phase discontinuity, also called branch cut. This can be seen in the panel in Fig.~\ref{fig:Gauss_Symm}(b.) for a single Zero-Pole pair. In our example, the complex Zero and Pole values are purposely chosen such that the branch cut is crossing the real axis. The sufficient condition for a metasurface to support a $2\pi$ resonant phase accumulation as a function of the real frequency is thus to possess a Zero-Pole pair separated by the real axis, as illustrated in Fig.~\ref{fig:Gauss_Symm}(c.). After unwrapping the phase discontinuity resulting from crossing the branch cut at $\omega = 1$, we obtain a $2\pi$ phase difference $\Delta \text{Arg}\left(F\right) = \text{Arg}\left(F(\omega_2)\right) - \text{Arg}\left(F(\omega_1)\right)$ as $\omega_1 \rightarrow 0$ and $\omega_2 \rightarrow 2$ (see Fig.~\ref{fig:Gauss_Symm}(c)). Cauchy's residue theorem is used to link the $2\pi$ phase accumulation on the real axis with the existence of a phase singularity in the upper part of the complex plane. To integrate  $\Delta \text{Arg}\left(F\right) = \int_{\omega_1}^{\omega_2}\frac{d \text{Arg}\left(F(\omega)\right)}{d\omega} d\omega$ in the complex plane, we consider the integral along a contour $C_l$ containing one singularity of charge $q$ such as the one displayed in Fig.~\ref{fig:Gauss_Symm}(b.). It consists of the line segment $[\omega_1,\omega_2]$ closed by a semi-circle $C_{sc}$ of radius R in the upper part of the complex plane. As a consequence, $\Delta \text{Arg}\left(F\right) = 2\pi q -\int_{C_{sc}}\frac{d \text{Arg}\left(F(\omega)\right)}{d\omega} d\omega$. However, it can be shown that $\int_{C_{sc}}\frac{d \text{Arg}\left(F(\omega)\right)}{d\omega} d\omega \rightarrow 0$ when the radius R increases, see SI for the exact case of the transmission coefficient. Consequently, $\Delta \text{Arg}\left(F\right) \rightarrow 2\pi q$ directly linking the phase accumulation on the real axis to the existence of a phase singularity above the real axis.

\begin{widetext}
\begin{center}
\begin{figure}[h!]
\includegraphics[width=0.8\textwidth]{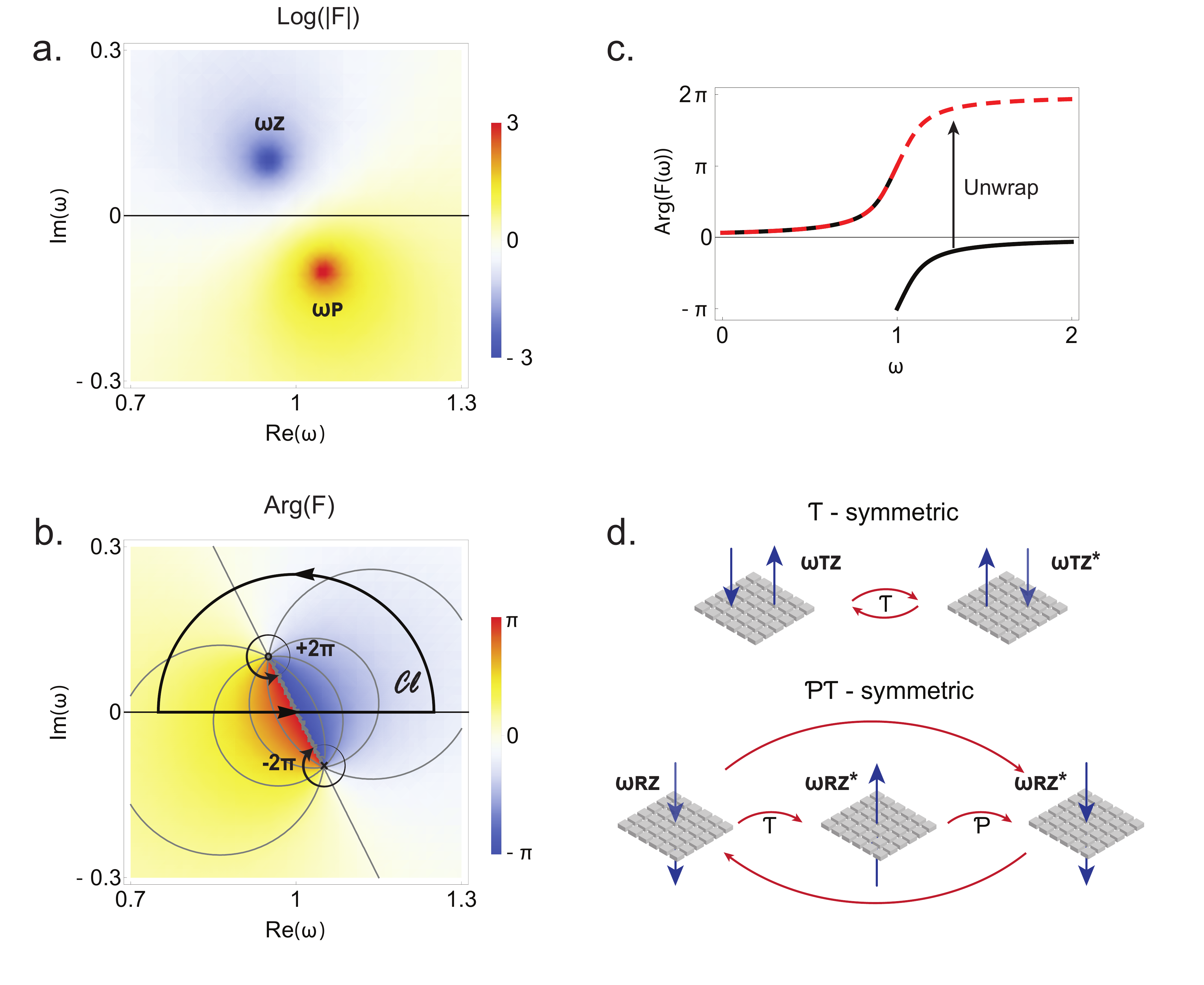}
\caption{\small \textbf{: Illustration of $2\pi$-phase accumulation concept relying on symmetry-breaking to position Zero and Pole across the real axis}. \textbf{(a.)} Logarithm map of a simplified model of metasurface optical response $F(\omega)=(\omega-\omega_z)/(\omega-\omega_p)$ as a function of the real and imaginary parts of the complex frequency, with $\omega_z$, and $\omega_p$, chosen so as to be in the upper part, and respectively in the lower part, of the complex plane. \textbf{(b.)} Argument of $F(\omega)$ in (a.) as a function of the real and imaginary part of the complex frequency featuring phase singularities at the positions of the Zero and the Pole of $F(\omega)$. Overlay black contour $C_l$ encircling the Zero used for the complex plane integration. \textbf{(c.)} Wrapped and unwrapped phase of $F(\omega)$ as a function of the (real) frequency. The discontinuity at $\omega = 1$ results from crossing the branch cut observed in (b.). \textbf{(d.)} Schematics of metasurfaces operating in reflection and respectively in transmission, supporting time-reversal ($\mathnormal{T}$) Transmissionless state at frequency $\omega_{\mathnormal{TZ}}$ and respectively Parity-Time ($\mathnormal{PT}$) Reflectionless state at frequency $\omega_{\mathnormal{RZ}}$.}
\label{fig:Gauss_Symm}
\end{figure}
\end{center}
\end{widetext}

\section{\label{sec:leve2} The link between symmetries and positions of zero-point singularities in the complex plane}
As a consequence of causality, Poles are always restricted to the lower part of the complex plane in passive systems \cite{nussenzveig1972} (the convention used for the time dependence throughout this article is $e^{-i\omega t}$). It is known that in active systems, adding gain can in principle move Poles upwards, fulfilling thereby the lasing condition when they reach the real axis, but this is not a simple and technologically relevant solution for metasurfaces. This is, therefore, outside of the scope of the present study. A question then arises: how to design a passive structure that can support a phase singularity in the upper part of the complex plane thus displaying a $2\pi$ resonant phase accumulation over the real axis? Discarding active systems with complex Pole manipulation, we propose to move Zeros to the upper part of the complex plane for the reflection or the transmission coefficients. Let us point out that for lossless structures or at least for structures featuring a small amount of loss, the Zero-Pole pairs of the S-matrix coefficients are usually separated by the real axis. Therefore, the S-matrix coefficients display a $2\pi$ phase accumulation for each of its Zero-Pole pairs. This can be seen in \cite{grigoriev_singular_2013} for a plasmonic dolmen metasurface. This is however of little practical interest for designing metasurfaces since the S-matrix links purely incoming fields to purely outgoing fields that are difficult to produce experimentally. In the remaining of this article, we will consequently focus on ways to move transmission and reflection Zeros to the upper half of the complex plane.  To this end, one first needs to better understand the constraints imposed on the locations of the Zeros in the complex plane by the symmetries of the metasurface.\\
\textbf{Transmissionless states of a system supporting Time-reversal symmetry}. 
Any dielectric metasurface made of lossless material defines a system which is unitary. A Transmissionless state  characterized by a Zero in the complex plane of the transmission coefficient thus imposes unity reflection  at this complex frequency. As applying time-reversal symmetry requires taking the conjugate of the operation frequency, any metasurface invariant under time-reversal possessing a Zero of transmission at $\omega_{\text{TZ}}$, should also possess another Zero of transmission at $\omega_{\text{TZ}}^{*}$ \cite{kang2021}. And the same is true for Reflectionless states.\\ 
\textbf{Reflectionless states of a system supporting Parity-Time symmetry}. \\
The study of the underlying symmetry properties of Reflectionless states necessitates considering the reflection symmetry with respect to the median plane (z=0) of the metasurface in addition to time reversal \cite{sweeney_theory_2020}. Let us consider a metasurface possessing a reflection zero for light impinging from one given side at a possibly complex-valued frequency $\omega_{\text{RZ}}$. If the metasurface is invariant under time-reversal, then it also possesses a reflection zero at $\omega_{\text{RZ}}^{*}$ for light impinging on the other side as illustrated in Fig.~\ref{fig:Gauss_Symm}(d.). In addition, the metasurface may remain invariant under the symmetry transformation which consists of a reflection with respect to its median plane. For the latter condition to be fulfilled, not only should the permittivity of the substrate $\epsilon_{\text{sub}}$ be equal to the permittivity of the top medium $\epsilon_{\text{sup}}$ but the  scatterers composing the metasurface should all remain invariant under this reflection transformation as well (which is the case for the nanocubes considered in the following of the article). A metasurface invariant under time reversal and under the latter reflection symmetry (that will be called parity symmetry in the following) possesses a reflection zero at both frequencies $\omega_{\text{RZ}}$ and $\omega^{*}_{\text{RZ}}$  for light impinging from the same side as shown Fig.~\ref{fig:Gauss_Symm}(d.).\\
\textbf{Real versus Complex-conjugate Zeros of symmetry-preserving systems}. These considerations indicate that the time-reversed counterpart of a Transmissionless state is also a Transmissionless state, and similarly, the parity-time symmetric of a Reflectionless state is also a Reflectionless state, as illustrated in Fig.~\ref{fig:Gauss_Symm}(d.). Upon conservation of symmetries, transmission or respectively reflection Zeros should thus occur either at real frequencies ($\omega_{\text{TZ}}$ is real) or as conjugated pairs in the complex plane ($i.e.$ $\omega_{\text{TZ}}$ and its complex-conjugated counterpart are located symmetrically with respect to the real axis). The former case is generic and could happen even for a single Zero-Pole pair, but to understand the latter case it is necessary to remind that  Eq.~\eqref{Weier_T} associates one Zero to only one Pole. Therefore, symmetry-preserving systems having conjugated Zeros positioned symmetrically with respect to the real axis imply necessarily the interaction of at least two Zero-Pole pairs. 

\section{Explicit or spontaneous symmetry-breaking leading to $2\pi$-phase accumulation on the real axis}
Given these symmetry constraints, two approaches can be considered to expel Zeros to the upper part of the complex plane:
\textbf{(i) Explicit breaking of a relevant symmetry}, which relaxes the real-value constraint for an isolated Zero, thus allowing to move it into the complex plane. This solution is quite straightforward and works both for reflection, and respectively transmission, depending if $\mathnormal{T}$-symmetry, or respectively $\mathnormal{PT}$-symmetry, breaking is engineered, for instance by introducing gain or losses or by putting the metasurface on a substrate. \textbf{(ii) Avoided crossing between two Zero-Pole pairs}: Considering two pairs, the system could remain $\mathnormal{T}$- or $\mathnormal{PT}$-symmetric as a whole, but decomposed into two states which are not symmetric anymore. Spontaneous symmetry-breaking of the Zero states thus creates the condition for which two different Zeros bifurcate as a conjugated pair in the complex plane. The bifurcation of the once-coalesced Zeros forces them to move symmetrically into the upper and lower halves of the complex plane. As the overall symmetries are preserved, the conjugated Zeros are positioned symmetrically with respect to the real axis. The trajectories of the Zero-Pole pairs as a whole, presented in the supplementary videos (without or with absorption losses), indicate that the trajectories of the two pairs are such that the branch cuts linking each pair avoid each other. To this end, one of the Zeros has to go down in the complex plane while the Pole from the other pair moves up so that the branch cuts from each pair do not cross. This will be further discussed in section \ref{sec_GSTC}
\footnotemark[4] \footnotetext[4]{While it is usual to look at the trajectories of resonance eigenfrequencies alone (crossing and anti-crossing), these results indicate that understanding of similar avoided crossing mechanisms in non-Hermitian systems requires to study the trajectories of the Zero-Pole pairs as a whole.}.

As a result, both scenarios leverage on Zeros expelled to the upper part of the complex plane to achieve a $2\pi$-phase accumulation caused by the branch cut of one zero-pole pair crossing the real axis. We illustrate both scenarios considering a metasurface composed of square nanocubes with a length $L=350$~nm and various heights $h$, made of a dielectric medium with relative permittivity $\varepsilon=8.05$ (which approximately corresponds to Sb$_2$S$_3$ phase-change material in the amorphous state) arranged in a 2D square array with a period $p=500$~nm. The substrate relative permittivity is equal to $\varepsilon_{\text{sub}} = 2.25$ while the relative permittivity of the top medium will be varied in the following examples. The numerical calculations of the optical response of these metasurfaces for complex-valued frequencies are performed using JCMsuite, a software based on the Finite-element method \cite{pomplun_adaptive_2007}.  For details on the numerical simulations, please see the Supporting Information section \ref{sec:level2}.  

\begin{widetext}
\begin{center}
\begin{figure}[h!]
\includegraphics[width=\textwidth]{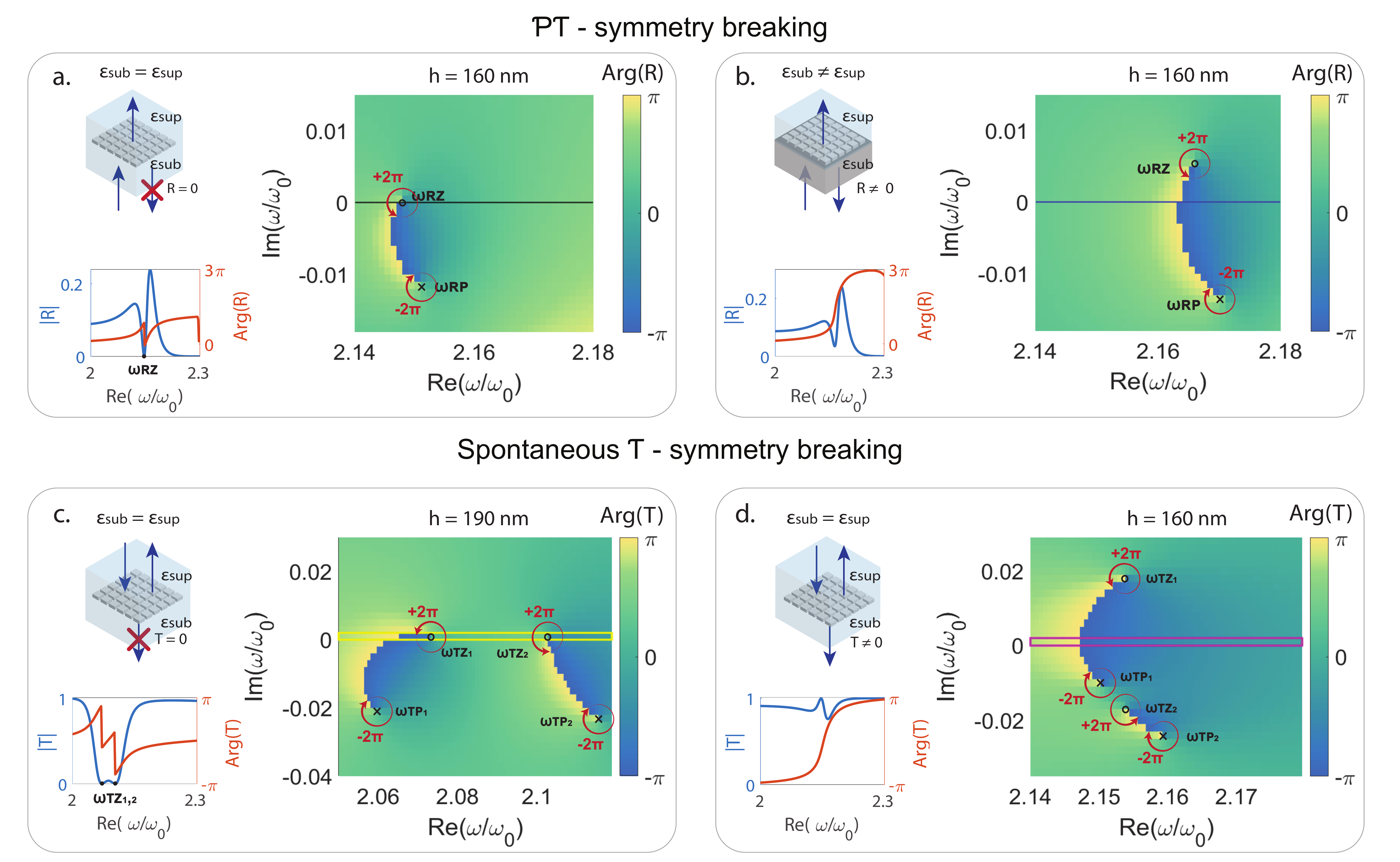}
\caption{\small \textbf{: Examples of two types of symmetry-breaking inducing a topological singularity to cross the real axis.} Each panel displays the $\text{Arg}(F)$ as a function of complex frequency $\omega$, computed by using numerical simulations, insets show the behavior for real frequency and a schematics of the setup. \textbf{(a., b.) Explicit $\mathnormal{PT}$-symmetry breaking. (a.)} Parity-time symmetry is preserved by surrounding the metasurface with homogeneous material. The Zero sits on the real axis and the accumulated phase is smaller than $2\pi$. \textbf{(b.)} Breaking parity symmetry is realized by modifying the refractive index of the superstrate ($\varepsilon_{\text{sup}} = 2.12$) which results in displacing the Zero to the upper part of the complex plane, leading to more than $2\pi$ phase accumulation in reflection. \textbf{(c., d.): Spontaneous $\mathnormal{T}$-symmetry breaking} \textbf{(c.)} Two Transmissionless states are present on the real axis. As a result, the transmission is modulated from 0 to 1, and poor phase coverage is achieved. \textbf{(d.)} The system is then driven to the condition known as Huygens metasurfaces by changing the height $h$ of the building blocks from $h = 190$ nm in (c.) to $h = 160$ nm in (d.). One of the Zeros of the two pairs is expelled in the upper complex plane. As a result, the transmission reaches almost unity and the phase accumulates up to $2\pi$. Spontaneous symmetry breaking happens after the coalescence of Zeros at EP. This process is explained in detail in Fig. \ref{fig:EP}. Note that \textbf{(d.)} structural parameters are strictly identical as in (a.), indicating that the high transmission window of HMS results from the existence of an associated $\omega_{RZ}$. The yellow and pink rectangles in (c.) and (d.) are indicating the respective regions denoted by the yellow and pink rectangles in Figs~\ref{fig:EP} (a.) and (b.).} 
\label{fig:Sym_break}
\end{figure}
\end{center}
\end{widetext}

\subsection{Explicit Parity-time symmetry-breaking for phase engineering in reflection}
To illustrate the first method, we start by considering a PT-symmetric version of the Sb$_2$S$_3$ metasurface: it is lossless and the superstrate is a dielectric material with the same relative permittivity as that of the substrate: $\varepsilon_{\text{sub}}=\varepsilon_{\text{sub}}=2.25$, making the overall system invariant under reflection with respect to its median plane at $z=0$. For nanocubes of height $160$ nm, it can be seen in Fig. \ref{fig:Sym_break} (a.) that a reflection Zero exists on the real axis for $\omega \approx 2.15\times10^{15}~\text{rad}\cdot\text{s}^{-1}$. The parity symmetry may be broken simply by changing the dielectric permittivity of the top material to $\varepsilon_{\text{sup}}=2.12$, keeping the substrate permittivity unchanged and preserving the time-reversal symmetry. The results for an incident field impinging from the substrate side are shown in Fig. \ref{fig:Sym_break} (b.). As a consequence of the parity symmetry breaking, the reflection Zero moves upwards. A drastic change of the behavior of the reflection coefficient phase in the $\omega$-plane can thus be observed. The plot of the phase dependency for real frequencies in Fig. \ref{fig:Sym_break} (a.) shows that there is a phase discontinuity when the Zero is located on the real axis. On the other hand, the phase varies from 0 to beyond $2\pi$ when the Zero moves to the upper part of the complex plane as seen in Fig.~\ref{fig:Sym_break} (b.). The reason why the phase accumulation exceeds $2\pi$ is the existence of another Zero-Pole pair in the lower part of the complex plane for larger frequencies (see SI). Transposing this to the transmission case, i.e. moving up the Transmissionless states, requires breaking the time-reversal symmetry by adding loss or gain in the metasurface \footnotemark[5]\footnotetext[5]{Note that to observe the resonant $2\pi$ phase accumulation, Pole and Zero still have to be positioned across the real axis. It means that the amount of added gain has to be carefully chosen for the  Pole to remain alone in the lower part of the complex plane. Note however that when the gain is added and poles are getting closer and closer to the real axis, nonlinear effects may become important. The description of a pole and zero positioning in the nonlinear regime is out of the scope of the present studies.}.

\subsection{Spontaneous symmetry-breaking at EP for phase engineering in transmission}
Let us now illustrate the second approach. The map of the phase of the transmission coefficient of the Sb$_2$S$_3$ metasurface for nanocube height equal to $190$~nm is displayed in Fig.~\ref{fig:Sym_break} (c.) showing the existence of two Zeros on the real axis. Decreasing the nanocube height to $160$~nm, qualitatively modifies the complex transmission map. The two Zeros leave the real axis and form a conjugated pair below and above the real axis, as shown in Fig.~\ref{fig:Sym_break} (d.). This is an example of spontaneous time-reversal symmetry breaking of the states \cite{lu2018}. The existence of a transmission Zero in the upper half of the complex plane provides a full $2\pi$ accumulation over the real axis. It is confirmed by looking at the phase accumulation as a function of the real frequency on the left bottom plot in Fig. \ref{fig:Sym_break} (d.). We emphasize that this approach would work in reflection, where a spontaneous breaking of the PT-symmetry would result in the appearance of a conjugated pair of reflection Zeros as well.

\subsection{Coalescence of Zeros at the exceptional points: the topological origin of the Huygens Metasurfaces}
The origin of the spontaneous symmetry breaking observed in Fig. \ref{fig:Sym_break}~(d.) requires the interaction between several Zero-Pole pairs. To study this interaction, the height of the considered metasurface is gradually varied from $h= 120$~nm to $h= 200$~nm. The maps of the amplitude and phase of the metasurface as a function of the height of the metasurface nanocubes and frequency are shown in Fig.~\ref{fig:EP} (a.) and (b.). These maps reveal a behavior identical to the one reported for dielectric Huygens metasurfaces (HMS) \cite{decker_highefficiency_2015} with the existence of a range of heights for which the amplitude of the transmission coefficient becomes large and the phase covers the full $2\pi$ interval. The real and imaginary parts of transmission Zeros as a function of the height of the metasurface nanocubes are shown in Fig.~\ref{fig:EP} (b.) and (c.). Outside the HMS operating region, the two Zeros occur for two different real parts but share the same imaginary part (which vanishes). However, the two Zero real parts converge close to the boundaries of the HMS operating region and their real parts become identical at the boundaries of this region. Their imaginary parts split into two values that are symmetric with respect to the real axis. Analyzing the trajectories of the transmission Zeros thus allows us to link the operating region of HMS occurring between $h\approx 142$~nm and $h \approx 183$~nm to the range of heights where transmission Zeros exist as a conjugated pair in the complex plane, which corresponds to the regime of spontaneously broken time-reversal symmetry and to the existence of two transmission Zeros degeneracies, also called exceptional points (EPs)\cite{sweeney_perfectly_2019}, at the boundaries of the HMS operating region. Avoided crossing of two Zero-Pole pairs moves one of the Zeros in the upper part of the complex plane satisfying the sufficient branch cut crossing condition required to achieve $2\pi$ phase coverage. The EPs are found for $h_{\text{EP1}} =183.58$~nm and $h_{\text{EP2}} = 142.88$~nm. EPs are a type of degeneracies specific to non-Hermitian systems \cite{heiss2012,zhen2015}. At the degeneracy, not only do the eigenfrequencies become identical but their associated eigenstates coalesce. In the field of metasurfaces, the coalescence of eigenstates at an EP has been used for the design of polarization-dependent metasurfaces recently \cite{song_plasmonic_2021}. EP have also been proven useful in many other applications \cite{miri2019}, in particular for sensing \cite{wiersig2014,park2020}. Figs.~\ref{fig:EP}~(d.) represents the phase of the transmission coefficient in the complex frequency plane around EP1. The circulation of the phase around the Zero indicates that it is a second-order Zero with phase winding equal to $4\pi$ leading to a topological charge equal to $q=+2$, as expected for an EP. To illustrate how the trajectories of Zeros are linked to the existence of these two EPs, we fit these trajectories with a superposition of a linear dependency with $h$ and a square root function with two Zeros thus featuring two square-root branch points at the positions of the two EPs:  
\begin{equation}
    \omega_{\text{TZ},EP1/EP2}(h) = a~h+b \pm c\sqrt{\left(h_{EP1}-h\right)\left(h_{EP2}-h\right)},
\end{equation}
the constants a, b and c are determined by fitting leading to the following values: $a = -2.3465 \times10^{21}~\text{s}^{-1}\cdot\text{m}^{-1}$, $b = 2.5317\times10^{15}~\text{s}^{-1}$ and $c=8.462 \times 10^{20}~\text{s}^{-1}\cdot\text{m}^{-1}$. The predictions of this fitting model are shown by red and black lines in Figs.~\ref{fig:EP} (b.) and (c.). A good qualitative agreement between the trajectories of zeros and the fitting function is found for both the real and imaginary parts. For height smaller than $130$~nm, a deviation between the fitting model $\omega_{\text{TZ},-}(h)$ and the position of Zeros determined by the numerical calculations (black dots and line) is observed. EPs are, by definition, square-root branch point singularities corresponding to conditions for which the expression under the square-root sign vanishes. This confirms that spontaneous symmetry breaking occurs at exceptional points of Transmissionless states. Here only the height of the nanocubes is tuned to obtain these EPs, while these are usually found by tuning two parameters. This is a consequence of the underlying symmetries of the system as pointed out in \cite{fruchart2021}. An important consequence of the discovery of this mechanism is that it provides a fundamental explanation of the physics underlying the design of widely used Dielectric Huygens metasurfaces (HMS) \cite{cheng_wave_2014, yu_high-transmission_2015, wang_grayscale_2016, li_polarization-independent_2016, zhao_dielectric_2016, shanei_dielectric_2017, zhou_efficient_2017, bar-david_situ_2017, tian_all-dielectric_2017, liu_huygens_nodate, ollanik_high-efficiency_2018, yoon_pragmatic_2018, iyer_reconfigurable_2015,komar_electrically_2017,li_phase-only_2019,leitis_alldielectric_2020, staude_tailoring_2013, decker_highefficiency_2015}. The original understanding of HMS properties relied exclusively on a spectral overlap of electric and magnetic dipolar resonances, which was derived following a theoretical modal analysis of arrays of spherical silicon particles \cite{evlyukhin_optical_2010}. By carefully studying the trajectories of Zeros, we prove that the boundaries of the $2\pi$ phase change regions correspond to EPs where two Transmissionless Zeros become degenerated. We also computed the electric and magnetic resonance frequencies of this Sb$_2$S$_3$ metasurface, and the silicon nanodisk metasurface proposed in \cite{decker_highefficiency_2015} (see SI), and we found that they never fully coincide. The $2\pi$ phase accumulation of the Huygens metasurfaces, originally attributed to the superposition of $\pi$ phase of two coinciding resonances \cite{decker_highefficiency_2015}, is the consequence of the discovered $2\pi$-phase mechanism reported herein, i.e. only one Zero-Pole branch-cut crossing the real axis.\\
Finally, it is worth mentioning that the symmetry considerations used throughout this article were determined for a field impinging on the metasurface at normal incidence. At non-normal incidence angles, additional spatial symmetries of the metasurface play an important role, in particular in-plane symmetries, explaining the high sensitivity of Huygens metasurfaces response with respect to the incident angle \cite{gigli_fundamental_2021}.

\begin{widetext}
\begin{center}
\begin{figure}[h!]
\includegraphics[width=\textwidth]{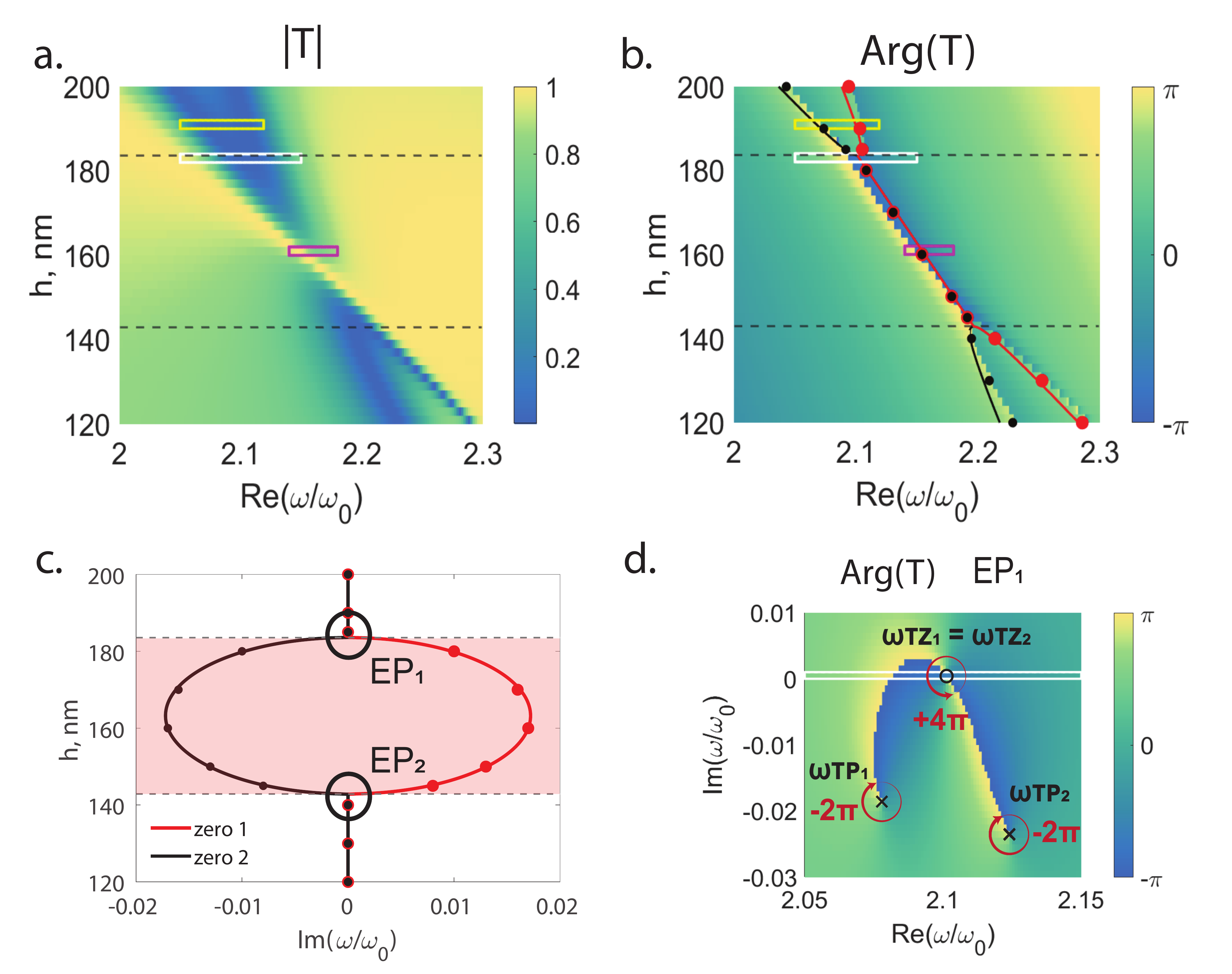}
\caption{\small \textbf{Huygens metasurfaces: a Spontaneous $\mathnormal{T}$-symmetry breaking condition.} \textbf{(a.)} Typical transmission spectra of metasurface operating in the Huygens regime as a function of the nanocube height $h$ and real part of the frequency. Two Transmissionless states are present in the top and bottom part of the graph, denoted by the dashed black lines. In the middle, a high transmission window opens, with near-unity transmission region indicated by the pink rectangle. \textbf{(b.)} Phase map associated to the transmission map in (a.). Superposed is shown the real parts of the frequencies of the Transmissionless states. They merge in two points, at the boundary of the Huygens regime, and remain with the same real frequency in-between. \textbf{(c.)} Imaginary parts of the Zeros as a function of the nanocube height $h$. At the boundary of the Huygens regime, the imaginary part of the Zero frequencies bifurcates and symmetrically move respectively to the upper and lower parts of the complex plane as a consequence of spontaneous symmetry breaking. \textbf{(d.)} Phase map as a function of the real and imaginary part of the frequency at the coalescence of Zeros $h_{\text{EP1}}=183.58$~nm, or EP. At the EP, the circulation of the phase is equal to $4\pi$. The white rectangle denotes the  region highlighted by the white rectangle in Figs~\ref{fig:EP} (a.) and (b.)}
\label{fig:EP}
\end{figure}
\end{center}
\end{widetext}
\section{\label{sec_GSTC}Electromagnetic susceptibility responses of topological surfaces and influence of absorption losses}

\begin{figure*}[h!]
\includegraphics[width=\textwidth]{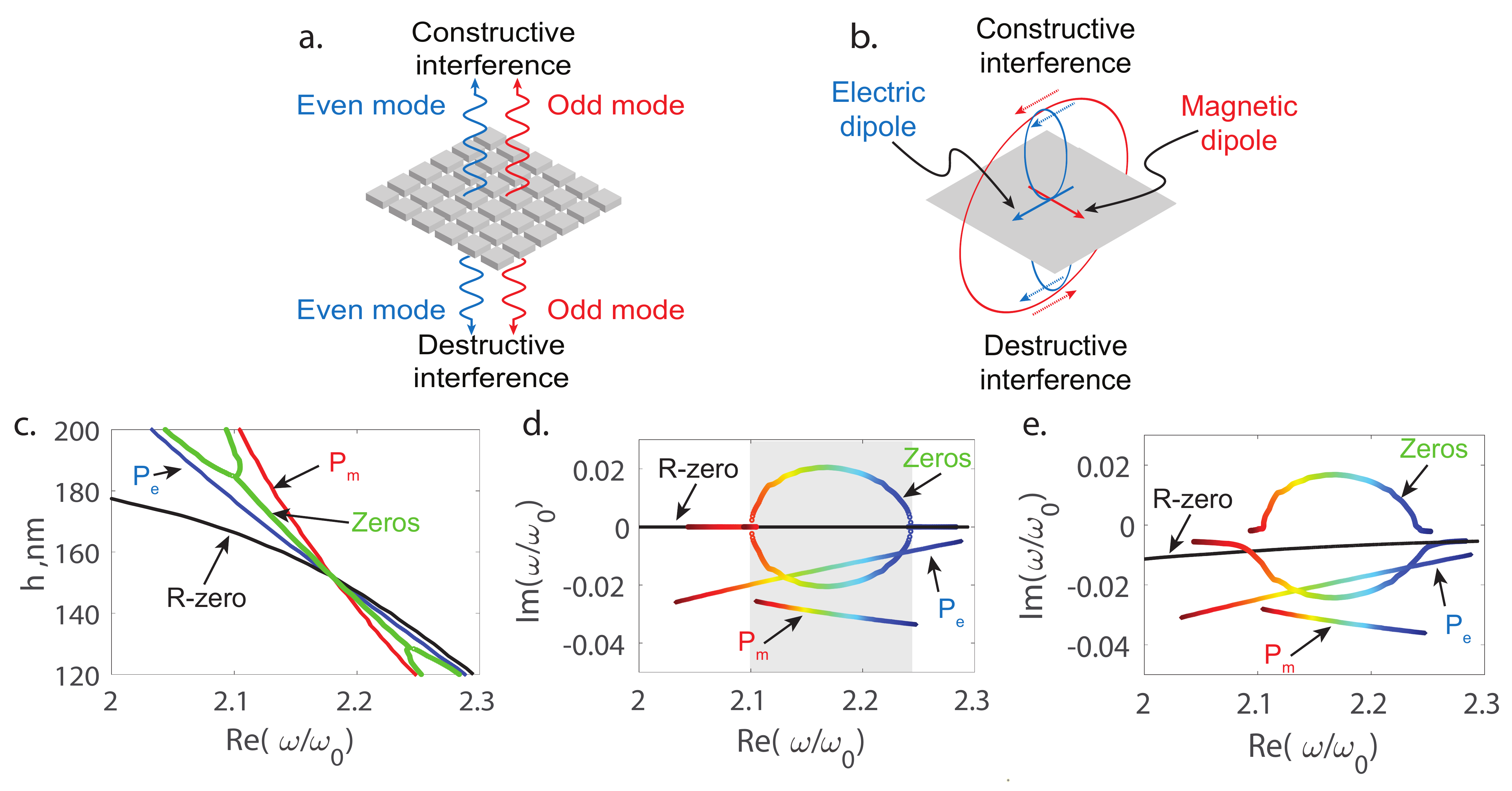}
\caption{\small \textbf{: Symmetries of the electric and magnetic modes and avoided crossing in the presence of intrinsic loss.} \textbf{(a.)} Schematic representation of even and odd mode interference at a metasurface presenting unidirectional scattering properties. \textbf{(b.)} Classical representation of Huygens E and M dipolar interference condition achieving unidirectional scattering in (a.). \textbf{(c.)} Real part of the frequency solutions of GSTCs for Zeros, reflection zero (denoted R-zero) and Poles ($\text{P}_\text{e}$ and $\text{P}_\text{m}$) positions as a function of the metasurface height. Avoided crossing of pairs occurs at the frequency where all curves intersect around $Re (\omega)\approx 2.17 \times  10^{15} (\text{rad.s}^{-1})$. \textbf{(d.)} Complex-plane positioning of the reflection zero (R-zero), transmission Zeros, and E and M-Pole frequencies as a function of the nanocube height (color-coded from blue to red for increasing height). Avoided crossing of the zero-pole pairs occurs such that the ``fastest" Pole associated with the electric resonances could cross the Zero-Pole branch cut of the magnetic mode. \textbf{(e.)} Same as in (d.) but in the presence of intrinsic losses, i.e., in the regime where explicit symmetry breaking of the time-reversal symmetry is realized. Zeros are expelled from the real axis, but for a similar topological reason as in (d.), avoided crossing of the Zero-Pole branch cut (previously found for time-reversal symmetric system) still occur, this case without degeneracy of Zeros at EPs.}
\label{fig:phase}
\end{figure*}

We started in Sec. II with a single Zero-Pole pair toy-model, then in Sec. IV a numerical calculation for a realistic metasurface with several Zero-Pole pairs and now, in Sec. V, we rely on another description (GSTC) of the metasurface response to specifically investigate the trajectories of transmission Zeros. GSTCs are convenient analytical tools to study the metasurface optical responses. They consist of a zero-thickness sheet supporting electric and magnetic dipole responses expressed in terms of effective surface susceptibilities~\cite{achouri2020electromagnetic}. In this way, we will be able to link our previous results obtained using symmetry-based discussions to the electric and magnetic dipole responses of metasurfaces which are widely used as a design tool. The effective metasurface susceptibilities provide simple expressions for the reflection $R$ or transmission $T$ coefficients. For a normally propagating incident wave, they read as~\cite{achouri2020electromagnetic}:
\begin{subequations}
\label{eq:MS_RT}
\begin{align}
R &= \frac{2ik(\chi_\text{e} - \chi_\text{m})}{(2-ik\chi_\text{m})(2-ik\chi_\text{e})},\label{eq:MS_RT1}\\
T &= \frac{4+\chi_\text{m}\chi_\text{e}k^2}{(2-ik\chi_\text{m})(2-ik\chi_\text{e})},\label{eq:MS_RT2}
\end{align}
\end{subequations}
where $\chi_\text{e}$ and $\chi_\text{m}$ are the metasurface electric and magnetic isotropic susceptibilities and $k=\omega/c$ with $c$ being the speed of light in the metasurface surrounding medium. It is important to notice that magnetic currents generate odd modes and electric currents generate even modes as seen in Fig. \ref{fig:phase}(a). Here, even and odd are related to the parity of the solutions with respect to the median plane of the structure $z = 0$. Because of modal symmetry, previous works on HMS rightfully explained that destructive interference between these modes on one or the other side of the metasurface may lead to the appearance of reflection and transmission Zeros, also related to the first and second Kerker conditions \cite{decker_highefficiency_2015}. Here, we particularly focus on Transmissionless states with zero optical backscattered signal, and show that the GSTC method provides insights on the trajectories of transmission Zeros observed in Fig. \ref{fig:EP}.\\
The GSTC method yields the expressions provided in Eqs. \eqref{eq:MS_RT1} and \eqref{eq:MS_RT2} for the reflection and transmission coefficients where  $\chi_\text{e}(\omega)$ and $\chi_\text{m}(\omega)$ are the electric and magnetic susceptibilities, represented by the Lorentzian functions:
\begin{equation}
\label{eq:LorentzX}
    \chi_\text{e/m}(\omega) = \frac{A_\text{e/m}}{(\omega_\text{e/m}^2 - \omega^2) - i\gamma_\text{e/m}\omega},
\end{equation}

where $A_\text{e/m}$ is the amplitude, $\omega_\text{e/m}$ is the resonance frequency and $\gamma_\text{e/m}$ is the damping factor. All these quantities are determined by fitting from numerical calculations. Reflection, transmission Zeros, as well as resonances, may then be found by solving Eqs. \eqref{eq:MS_RT1} and \eqref{eq:MS_RT2}. Poles of $R$ and $T$ occur when either of the two following conditions are fulfilled independently for electric or magnetic resonances: $2-ik\chi_\text{e}=0$ and  $2-ik\chi_\text{m}=0$. However, Eqs. \eqref{eq:MS_RT1} and \eqref{eq:MS_RT2} indicate that reflection and respectively transmission Zeros occur when $\chi_\text{e} = \chi_\text{m}$ and respectively $4+\chi_\text{m}\chi_\text{e}k^2 = 0$. A precise mixed contribution of both electric and magnetic susceptibilities is therefore required to obtain transmission Zeros which result from the interferences between field distributions of different symmetries. Substituting Eq. \eqref{eq:LorentzX} into the different conditions is used to obtain the complex frequencies of transmission and reflection Zeros. The resulting expressions are provided in Eqs.~\eqref{eq:dampK} to~\eqref{eq:udampZ} of the Supplementary Material for the cases of damped ($\gamma_\text{e}\neq 0, \gamma_\text{m}\neq 0$) and undamped ($\gamma_\text{e}=\gamma_\text{m}= 0$) metasurfaces. Fig. \ref{fig:phase} (c) shows the variation of real parts of the transmission Zero, the reflection zero frequency and Pole frequencies as a function of the nanocubes height, $h$. The plot Fig. \ref{fig:phase} (d) and the additional Video 1 clearly highlight that the high transmission window, known as Huygens metasurface regime, results from Pole inversion through avoided crossing of two Zero-Pole pairs as the geometrical parameters are varied, creating an ``opening" near the real axis for the electric Zero-Pole pair to pass. From the physical point of view, avoided crossing of the pairs is imposed because the trajectories and velocities of the Poles displacements in the complex plane as a function of the height strongly differ due to different modal confinements of the electric and magnetic resonant fields. Poles and Zeros are being connected via the branch cut, i.e. the line where the phase is discontinuous, and therefore Pole inversion in the complex plane cannot occur directly by crossing branch cuts. The only topologically valid solution is to expel the two Zeros symmetrically from the real axis, so as to preserve $\mathnormal{PT}$-symmetry, in such a way that the ``fastest" Pole-Zero pair in terms of frequency shift as a function of geometrical parameters, could cross, to the other side of the ``slow" resonance. Due to larger field confinement, the electric resonance is more sensitive to structural parameter changes and it moves faster with respect to the magnetic mode. Repelling both Zeros from the real axis results in a high transmission window while one Zero located in the upper part of the complex plane introduces a $2\pi$-phase accumulation that is of interest for metasurface design. Fig.~\ref{fig:phase} (e) shows the same configuration as previously but with additional damping in the GSTC formula ($\gamma_\text{e}\neq 0, \gamma_\text{m}\neq 0$). In this case, time-reversal symmetry is broken. It can then be noted that there is no EPs of transmission Zeros. In fact, it is well known that the tuning of two parameters is usually necessary for reaching an EP. The reason why only one was necessary in the case of the lossless metasurface considered previously was the underlying symmetry of the system \cite{fruchart2021}, in our case time-reversal symmetry. As soon as this symmetry is broken, two parameters are again needed for reaching an EP. In Fig.~\ref{fig:phase} (e), the Zeros thus do not merge, but the avoided crossing behavior remains, leading again to the existence of one Zero in the upper half of the complex plane. As a result, the $2\pi$-phase accumulation is still observed without Zeros degeneracy and EP, indicating that this approach is robust and would also apply to metasurfaces made of arbitrary materials.

\section{\label{sec:conclusion}Conclusion}

In summary, we have proposed a general method to address $2\pi$ phase accumulation of the reflected or transmitted light at interfaces. We discovered that a sufficient condition is to have a system presenting a pair of topological phase singularities with opposite charges, the so-called Zero-Pole pair, located in the complex plane with the two singularities of the same pair located on each side of the real axis. In this condition, the branch cut crosses the real axis, thus accumulating a $2\pi$ phase as a function of the real frequency. Relying on symmetry arguments, two simple solutions to achieve such Zero-Pole separation are provided. The first solution consists of explicitly breaking the symmetry of the system, either breaking $\mathnormal{T}$-symmetry by adding gain or losses or $\mathnormal{PT}$-symmetry breaking by disposing the metasurface at the interface between two optically different materials. The second approach relies on the avoided crossing of two different Zero-Pole pairs. It is shown to induce a spontaneous symmetry-breaking that bifurcates coalesced Zeros at the EP, forcing them to symmetrically move out oppositely from the real axis to the upper and lower parts of the complex plane. This second approach suggests that topological arguments are at the origin of the physics underlying the design of Huygens metasurfaces. Harnessing degeneracies of several topological phase singularities and adding other symmetry-breaking conditions would open new design perspectives in nanophotonics. It reveals also that Huygens metasurfaces are a part of a much wider class of metasurfaces, greatly broadening the range of applicability of this design approach for controlling the phase of light with nanoscale topologically engineered symmetry-breaking structures.

\begin{acknowledgments}
PG and NB acknowledge financial support by the French National Research Agency ANR Project DILEMMA (ANR-20-CE09-0027). PG and EM acknowledge financial support by the French National Research Agency ANR Project Meta-On-Demand (ANR-20-CE24-0013). PG thanks J.Y. Duboz for useful discussions on the part related to complex plane integration. OM and KA acknowledge funding from the European Research Council (ERC-2015-AdG-695206 Nanofactory) and from the Swiss National Science Foundation (project PZ00P2\_193221). RC and SB acknowledge funding from the German Research Foundation (DFG, Excellence Cluster MATH+, EXC-2046/1, project 390685689), the
Helmholtz Association (Helmholtz Excellence Network SOLARMATH, project ExNet-0042- Phase-2-3).

\end{acknowledgments}


\bibliography{PRL_bib}

\newpage

\renewcommand{\thefigure}{S\arabic{figure}}

\setcounter{figure}{0}

\renewcommand{\theequation}{S\arabic{equation}}

\setcounter{equation}{0}

\renewcommand{\thesection}{S\Roman{section}}

\setcounter{section}{0}

\onecolumngrid
\begin{center}
\section*{Supplemental Material}
\end{center}

\section{\label{sec:Toy_model}Toy-model used in Fig. \ref{fig:Gauss_Symm}:}

In order to illustrate the impact of the relative positions of the Zero-Pole pair separated by the real axis, we used a toy model based on the Weierstrass expansion product provided in Eq. \eqref{Weier_T} of the main text. We used $A=1$ and $B=0$ and a single Zero-Pole pair: 
\begin{equation}
    F(\omega)=\frac{\omega-\omega_z}{\omega-\omega_p},
\end{equation}
where $\omega_z = 0.95 + i0.1$ and $\omega_p = 1.05 - i0.1$

\section{\label{sec:Top_charge}Topological charge and residue theorem:}

Let $F(\omega)$ be a response function characterizing the optical response of a metasurface. In general, $F(\omega)$ possesses a collection of Zeros $\omega_{z,i}$ and Poles $\omega_{p,i}$ located in the complex plane. Let us consider the logarithmic derivative of $F(\omega)$ : $\frac{d \log\left(F(\omega)\right)}{d\omega}$. This function possesses first-order Poles at the Zeros and Poles of $F(\omega)$. We can also easily compute the residue associated with these Poles, considering the case of zeros of $F(\omega)$ independently from the case of Poles of $F(\omega)$.\\
If $\omega_{z,i}$ is a Zero of order $q$ of $F(\omega)$, then $F(\omega)$ can be well approximated by the function $(\omega-\omega_{z,i})^qg(\omega)$, $g(\omega)$ being a regular function with no Zeros and Poles, for $\omega$ belonging to the vicinity of $\omega_{z,i}$. $log\left(F(\omega)\right)$ can then be approximated by the following expression for $\omega$ belonging to the vicinity of $\omega_{z,i}$:
\begin{equation}
\frac{d \log\left(F(\omega)\right)}{d\omega}\approx \frac{q}{\omega-\omega_{z,i}} + \frac{1}{g(\omega)}\frac{dg(\omega)}{d\omega},
\label{Eq:logF_approx}
\end{equation}
The residue of $\frac{d \log\left(F(\omega)\right)}{d\omega}$ associated with its Pole at $\omega_{z,i}$ is by definition equal to $\lim_{\omega\rightarrow \omega_{z,i}}(\omega-\omega_{z,i})\frac{d \log\left(F(\omega)\right)}{d\omega}$. Eq. \eqref{Eq:logF_approx} allows us to show that this residue is equal to $q$. If now, a Pole $\omega_{p,i}$ of $F(\omega)$ of order $q$, analogous derivations allow us to show that its residue for $\frac{d \log\left(F(\omega)\right)}{d\omega}$ is equal to $-q$.\\
Several results can be derived using Cauchy's residue theorem.\\
First of all let us consider a contour surrounding a single singularity $C$, $i.e.$ either a Zero or a Pole of $F(\omega)$. The previous results and the residue theorem allow us to show for an integration performed counterclockwise $\oint_{C} \frac{d \log\left(F(\omega)\right)}{d\omega} d\omega = 2i\pi q $ for a Zero of order $q$ of $F(\omega)$. On the other hand, $\oint_{C} \frac{d \log\left(F(\omega)\right)}{d\omega} d\omega = -2i\pi q $ for a Pole of order $q$ of $F(\omega)$. These results are used to compute the topological charge in the main text.\\
Now if we consider a contour containing several Zeros $\omega_{z,i}$ of order $q_{z,i}$ and Poles $\omega_{p,i}$ of order $q_{p,i}$, the residue theorem leads to the following result: 
\begin{equation}
\oint_{C} \frac{d \log\left(F(\omega)\right)}{d\omega} d\omega = 2i\pi\left(\sum_{\omega_{z,i}}q_{z,i}-\sum_{\omega_{p,i}}q_{p,i}\right).
\label{Eq:res_theorem}
\end{equation}

As a result, the contour integral on $\oint_{C} \frac{d \log \left(F(\omega)\right)}{d\omega} d\omega$ vanishes when $C$ contains both a simple Pole and a simple Zero for example. Eq. \eqref{Eq:res_theorem} is more general.\\

Let us emphasize that the Pole expansion of the logarithmic derivative of a response function is the starting point for deriving the Weierstrass product expansion provided in Eq. \eqref{Weier_T} (see supplemental material in \cite{grigoriev_optimization_2013}).

\section{Phase accumulation and Gauss theorem:}

Here, we are providing some additional derivations allowing us to illustrate the link between positions of topological charges and the phase accumulation on the real axis. To this end, we consider the type of contours shown in Fig. \ref{fig:Gauss_Symm} of the main text: a contour made of a segment on the real axis plus a semi-circle in the upper half of the complex plane.\\
We will use derivations for the transmission coefficient to illustrate these results. Since these derivations require determining the logarithmic derivative of the transmission coefficient, we will use here the analytical formulas provided by the GSTC method (see Eqs. \eqref{eq:MS_RT1} and \eqref{eq:LorentzX} of the main text). We will in particular consider the $Sb_2S_3$ metasurface considered in the Fig.~\ref{fig:Sym_break}(c.) and (d.) and Fig.~\ref{fig:EP} of the main text for a height $h=160$~nm.
The values of the parameters in the GSTC formulas for this case are: $A_e = 1.235\times10^{22}$ m.s$^2$, $\omega_e = 2.153\times10^{15}$ rad.s$^{-1}$, $\gamma_e = 0$, $A_m = 2.363\times10^{22}$ m.s$^2$, $\omega_m = 2.166\times10^{15}$ rad.s$^{-1}$ and $\gamma_m = 0$.\\

\begin{figure*}[h!]
    \centering
    \includegraphics[width=0.7\textwidth]{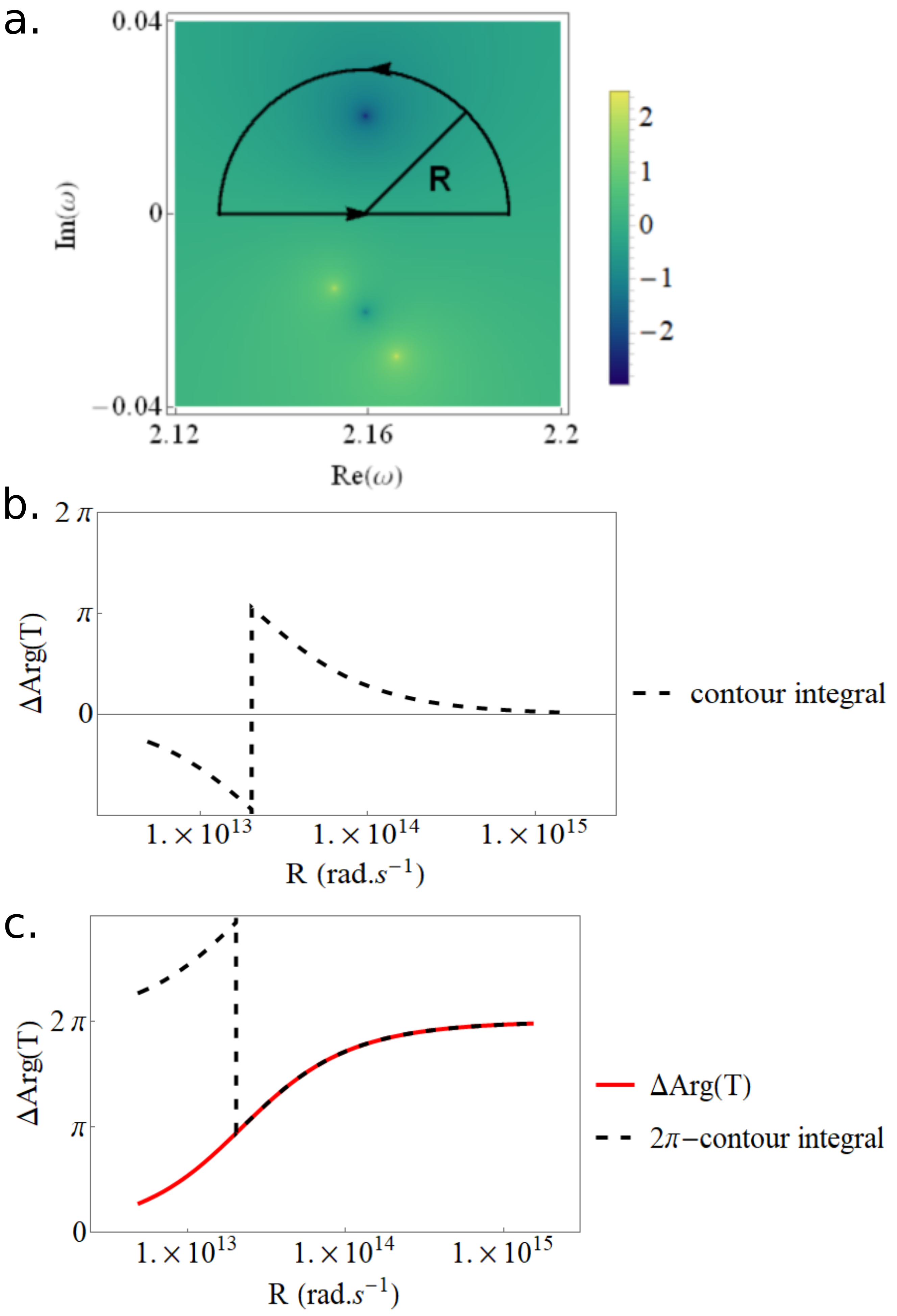}
    \caption{\small a. Map of the transmission coefficient in the complex frequency plane provided by the GSTC method for the $Sb_2S_3$ metasurface with $h=160$~nm $\omega_0 = 10^{-15}$ rad.s$^{-1}$. b. Dependency with $R$ of the semi-circle contour integral on the right hand side of Eq. \eqref{Eq:Phase_acc}. c. Dependency of $Arg(T)$ with $R$ showing a convergence towards $2\pi$ as R increases.}
    \label{fig:Gauss_theorem_Phase}
\end{figure*}

We want to compute the phase difference between two frequencies $\omega_2$ and $\omega_1$: $\Delta Arg(T) = Arg(T(\omega_2))- Arg(T(\omega_1))$. For the sake of simplicity, we choose $\omega_2$ and $\omega_1$ that are symmetric with respect to a center frequency $\omega_0$: $\omega_1 = \omega_0-R$ and $\omega_2 = \omega_0+R$. 
Computing this phase difference can be done by performing the following integration: $\Delta Arg(T) =\int_{\omega_0-R}^{\omega_0+R}\frac{d Arg(T)}{d\omega}d\omega = \text{Im}\left(\int_{\omega_0-R}^{\omega_0+R}\frac{d log(T)}{d\omega}d\omega\right)$. The evaluation of the latter integral will be done via a deformation of the contour in the complex plane.\\ 
Let us then consider the following integral: $\oint_{C} \frac{d\log\left(T(\omega)\right)}{d\omega} d\omega$ where $C$ is the contour shown in Fig.~\ref{fig:Gauss_theorem_Phase}(a.). Since there is only one Zero within this contour, this integral is equal to $2i\pi$ (see results in the previous section). The contour is split up into a segment on the real axis $[\omega_0-R;\omega_0+R]$, where $\omega_0 = 2.159\times10^{15}$ and $R$ is varied, and the semi circle of radius $R$ in the upper complex plane centered on $\omega_0$: 
\begin{equation}
\begin{aligned}
   \text{Im}\left(\oint_{C} \frac{d\log\left(T(\omega)\right)}{d\omega} d\omega\right) = 2\pi,\\
  \text{Im}\left(\int_{\omega_0-R}^{\omega_0+R}\frac{d\log\left(T(\omega)\right)}{d\omega} d\omega\right)+ \text{Im}\left(\int_{0}^{\pi}\frac{d\log\left(T(\omega_0+Re^{i\theta})\right)}{iRe^{i\theta}
   d\theta} iRe^{i\theta} d\theta\right)  =2\pi,
\label{Eq:Phase_acc}
\end{aligned}
\end{equation}
Consequently, in order to have $Arg(T)\rightarrow 2\pi$, we need to show that $\text{Im}\left(\int_{0}^{\pi}\frac{d\log\left(T(\omega_0+Re^{i\theta})\right)}{iRe^{i\theta}
   d\theta} iRe^{i\theta} d\theta\right)\rightarrow0$ as $R$ increases.\\
   The behavior of the semi-circle contour term on the right hand side of Eq.~\eqref{Eq:Phase_acc} as $R$ increases is shown in Fig.~\ref{fig:Gauss_theorem_Phase}(b.). The discontinuity corresponds the radius at which the Zero enters the contour. Then, for larger radii, we see clearly that this contour term decreases towards $0$ and eventually vanishes when $R$ is of the order of $10^{15}$. As a consequence, the integral on the real axis in the right-hand side of Eq. \eqref{Eq:Phase_acc} tends toward $2\pi$ as seen in Fig.~\ref{fig:Gauss_theorem_Phase}. This result allows to directly link the behavior of the phase on the real axis to the existence of a phase singularity in the upper half of the complex plane.

\section{\label{sec:level2}Real and complex frequency numerical calculation}
Results from Fig.~\ref{fig:Sym_break}(a.)-(d.) and Fig.~\ref{fig:EP}(d.) were obtained using the Maxwell solver JCMsuite based on a higher-order finite element method \cite{pomplun_adaptive_2007}. It allows calculating a complex transmission coefficient for complex excitation frequencies from which we extract transmission amplitude and phase.\\
The trajectories of the transmission Zeros shown in Fig.~\ref{fig:EP}(b.) and (c.) were determined by locating the Zeros in complex-frequency plane maps of the amplitude of the transmission coefficient.
The maps in Fig.~\ref{fig:EP}(a.) and (b.) are computed using CST Studio suite. Simulations from CST Studio suite are also used for determining the susceptibilities used for the GSTC method.

\section{Resonance calculations for the $Sb_2S_3$ and silicon Huygens metasurface:}

Here we show the trajectories of the electric and magnetic eigenfrequencies for the $Sb_2S_3$ metasurface used in the main text and the silicon metasurface proposed in \cite{decker_highefficiency_2015}. These results were derived by solving an eigenvalue problem where outgoing boundary conditions were imposed in the regions at the top and bottom of the metasurface. These calculations were performed using the eigenvalue solver implemented in JCMsuite \cite{binkowski_auxiliary_2019}.  
\begin{widetext}
\begin{center}
\begin{figure*}[h!]
\includegraphics[width=\textwidth]{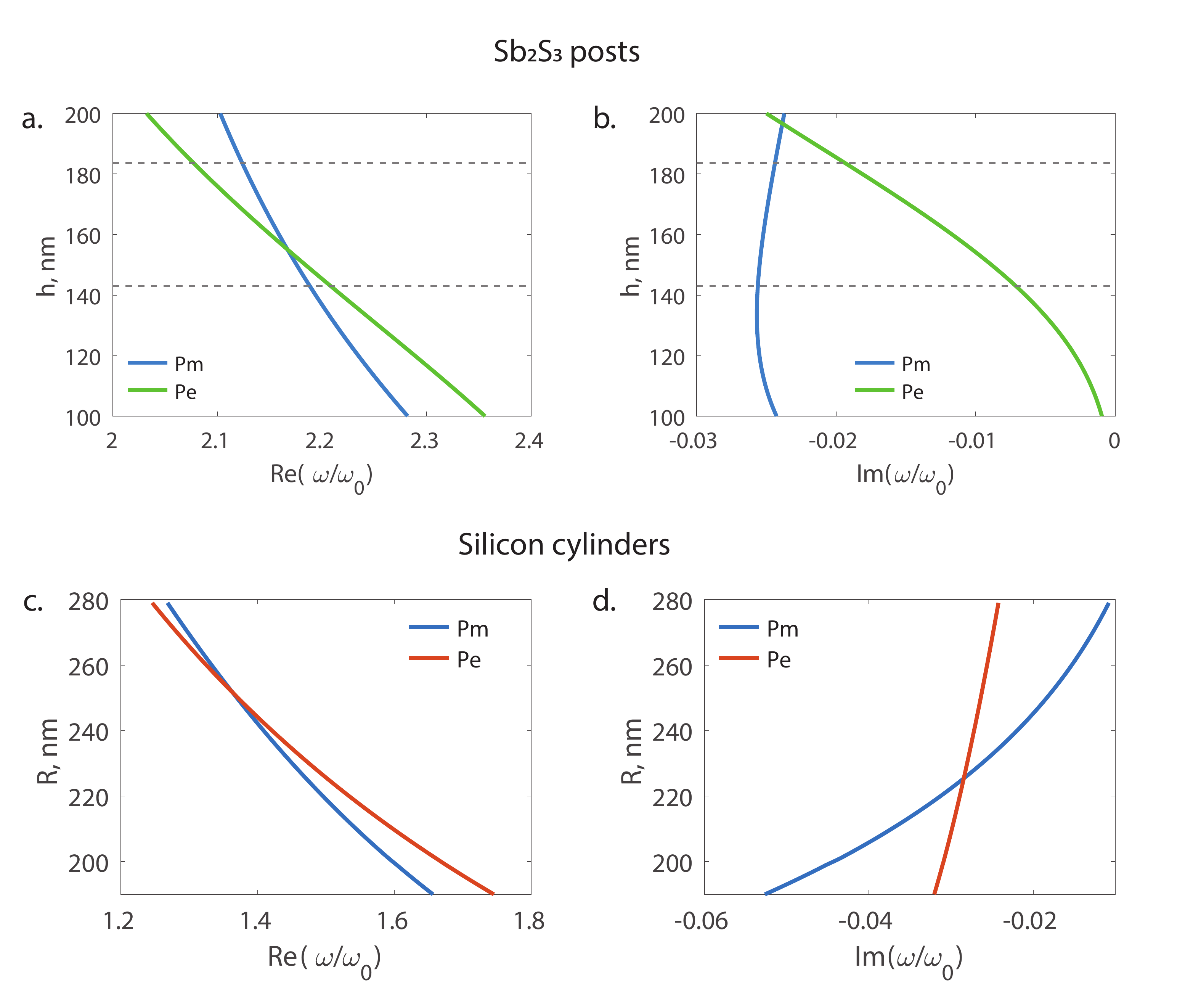}
\caption{\small a. Real part of the electric ($P_e$) and magnetic ($P_m$) resonance eigenfrequencies of the $Sb_2S_3$ metasurface for varying heights. b. Imaginary part of the resonance eigenfrequencies of the $Sb_2S_3$ metasurface. c. Real part of the electric ($P_e$) and magnetic ($P_m$) resonance eigenfrequencies of the metasurface made of silicon nanodisks studied in \cite{decker_highefficiency_2015} for varying radii. d. Imaginary part of the resonance eigenfrequencies of the silicon nanodisk metasurface.}
\label{fig:reson_SI_meta}
\end{figure*}
\end{center}
\end{widetext}
We note that, in both cases, the two Poles never fully coincide. For the $Sb_2S_3$ metasurface \ref{fig:reson_SI_meta} (a.) and (b.), there is a crossing of the real parts of the electric and magnetic resonance frequencies in the Huygens metasurface operating region (between the dashed lines) but their imaginary parts never cross. Similarly, the resonances of the silicon nanodisk metasurface never fully coincide as can be seen in Fig. \ref{fig:reson_SI_meta} (c.) and (d.).

\section{Additional results metasurface working in reflection:}

\subsection{Second Zero-Pole pair contributing to the phase accumulation in Fig.~\ref{fig:Sym_break} (b.)}
\begin{figure*}[h!]
    \centering
    \includegraphics[width=\textwidth]{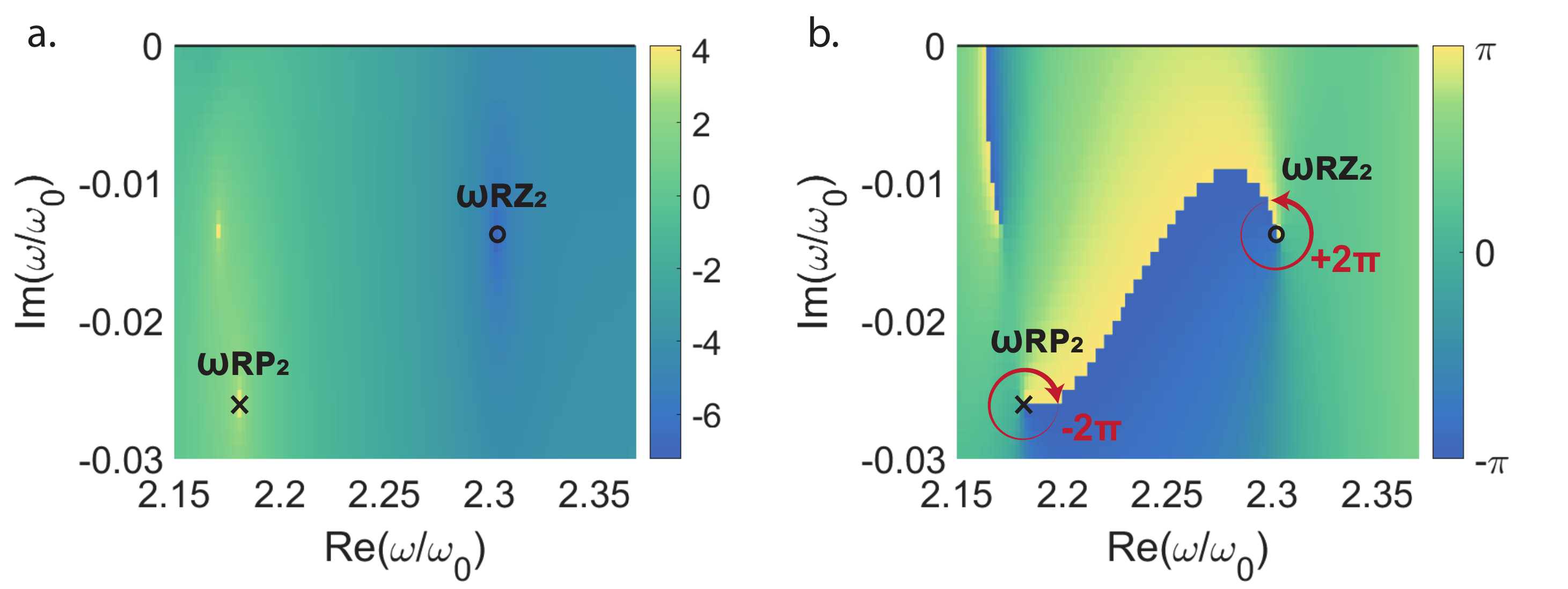}
    \caption{\small a. Complex-frequency map of $\log\lvert R(\omega)\rvert$ for the asymmetric structure studied in Fig.~\ref{fig:Sym_break} (b.) in the main text for larger frequencies $\omega_0 = 10^{15}$ rad.$s^{-1}$. b. Corresponding complex-frequency map of $\text{Arg}\left( R(\omega)\right)$}
    \label{fig:SI_second_zero}
\end{figure*}

In Figs \ref{fig:Sym_break} (a.) and (b.), it was shown how the explicit parity symmetry breaking could be used to design a metasurface displaying a $2\pi$ resonant phase accumulation over the real axis. It was in fact noted that the phase accumulation is larger than $2\pi$ in Fig.~\ref{fig:Sym_break} (b.). This is due to the existence of an additional Zero-Pole pair at larger frequencies as can be seen in Fig~\ref{fig:SI_second_zero}. Given that this second Zero remains under the real axis, it doesn't lead to an additional $2\pi$ accumulation on the real axis but it adds instead about $\pi$ to the phase accumulation seen in Fig.~\ref{fig:Sym_break} (b.). 

\subsection{Time-reversed reflection zero in the structure considered in Fig.~\ref{fig:Sym_break} (b.)}

\begin{figure*}[h!]
    \centering
    \includegraphics[width=\textwidth]{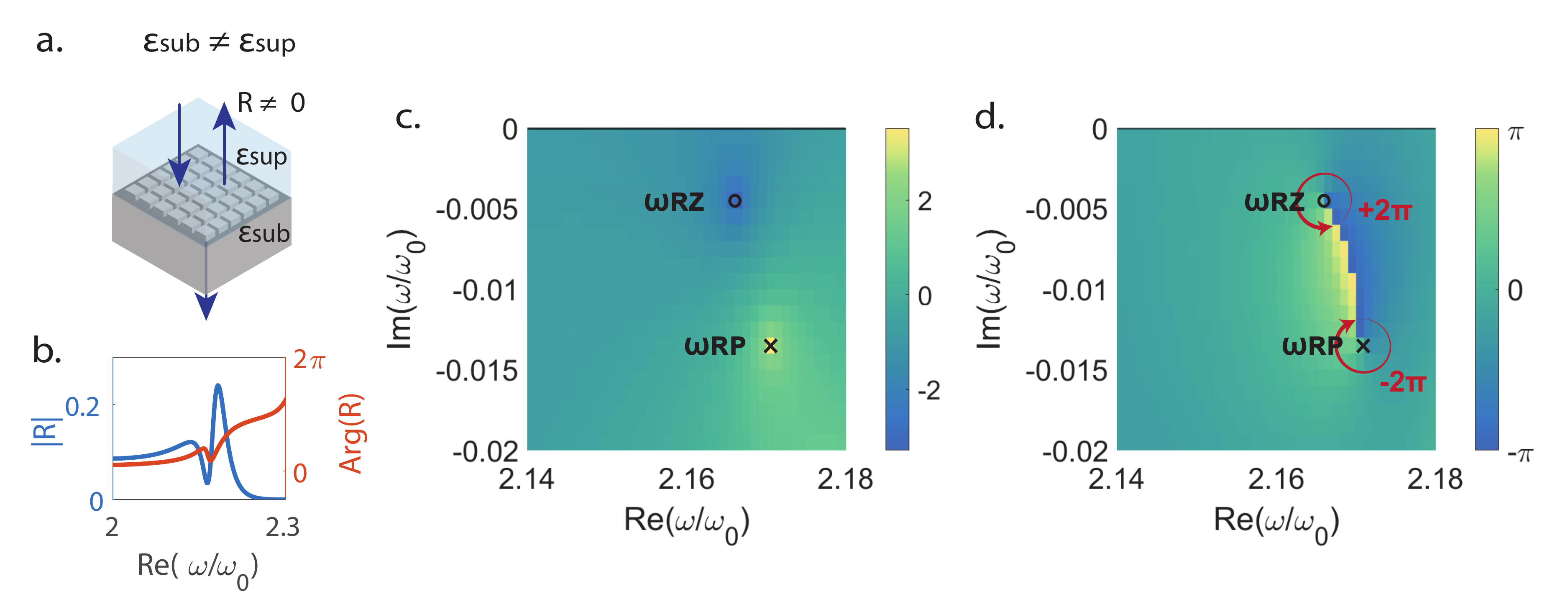}
    \caption{\small a. Optical response of the metasurface breaking the parity symmetry with $\epsilon_{\text{sup}}\neq\epsilon_{\text{sub}}$ for the same structure as in Fig.~\ref{fig:Sym_break} (b.) for light impinging from the top medium. b. Amplitude and phase of the reflection coefficient for real frequencies. c. Complex-frequency map of $\log\left(\lvert R(\omega)\rvert\right)$. d. Complex-frequency map of $\log\left(\lvert R(\omega)\rvert\right)$.}
    \label{fig:RZ_conjugate}
\end{figure*}

Fig.~\ref{fig:Sym_break} (b.) shows the existence of the reflection zero  for a frequency $\omega_{\text{RZ}}$ located in the upper half of the complex plane for an asymmetric structure with the permittivity of the substrate $\epsilon_{\text{sub}}$ different from the permittivity of the top medium $\epsilon_{\text{sup}}$ when light is impinging from the substrate. Time-reversal symmetry imposes the existence of a reflection zero at $\omega_{\text{RZ}}^{*}$ for light impinging from the top medium. In Fig.~\ref{fig:RZ_conjugate} (c.) and (d.), the complex frequency map of $\log\left(\lvert R(\omega)\rvert\right)$ and $\text{Arg}\left( R(\omega)\right)$ are shown for light impinging from the top medium. A reflection zero is seen at about $\omega/\omega_0 \approx 2.16 - i0.005$ which is indeed the conjugated value of the reflection zero observed in Fig.~\ref{fig:Sym_break} (b.). Since the zero remains in the lower half of the complex plane, the phase accumulation on the real axis is smaller than $2\pi$ as seen in Fig.~\ref{fig:RZ_conjugate} (b.). 

\section{Metasurface Susceptibility Modeling}

For known scattering parameters, relations~\eqref{eq:MS_RT} may be transformed to compute the metasurface effective susceptibilities as~\cite{achouri2020electromagnetic}
\begin{subequations}
\label{eq:X}
\begin{align}
    \chi_\text{e} &=- \frac{2i\left(T+R-1\right)}{k\left(T + R + 1\right)},\\
    \chi_\text{m} &=- \frac{2i\left(T-R-1\right)}{k\left(T - R + 1\right)}.
\end{align}
\end{subequations}

For the general case of a metasurface with damped Lorentzian responses, the complex frequencies  corresponding to the Kerker condition (i.e., full transmission through the metasurface, $R=0$) are obtained by using~\eqref{eq:LorentzX} within the Kerker condition on susceptibilities: $\chi_\text{e}(\omega) = \chi_\text{m}(\omega)$ and solving the resulting expression for $\omega$, which leads to
\begin{equation}
\label{eq:dampK}
        \omega_\text{K} = \frac{i}{2(A_\text{m}-A_\text{e})}\Big[A_\text{e}\gamma_\text{m}-A_\text{m}\gamma_\text{e}    \pm\sqrt{(A_\text{e}\gamma_\text{m}-A_\text{m}\gamma_\text{e})^2-4(A_\text{e}-A_\text{m})(A_\text{e}\omega_\text{m}^2-A_\text{m}\omega_\text{e}^2)}\Big].
\end{equation}
Following a similar procedure, the complex frequencies corresponding to the Poles are obtained from the resonance conditions : $2-ik\chi_\text{e/m}=0$. The results read as
\begin{equation}
\label{eq:dampP}
        \omega_\text{P} = -\frac{i}{4c}\left[A_\text{e/m}+2c\gamma_\text{e/m} \pm i\sqrt{16c^2\omega_\text{e/m}^2-(A_\text{e/m}+2c\gamma_\text{e/m})^2}\right].
\end{equation}
The analytical expression of the complex frequencies corresponding to transmission Zeros of a damped metasurface being very lengthy is omitted here.

In the case of an undamped metasurface (i.e., $\gamma_\text{e}=\gamma_\text{m}=0$), the relations~\eqref{eq:dampK} and~\eqref{eq:dampP} respectively reduce to
\begin{equation}
\label{eq:udampK}
        \omega_\text{K} = \pm\frac{\sqrt{A_\text{m}\omega_\text{e}^2 - A_\text{e}\omega_\text{m}^2}}{\sqrt{A_\text{m} - A_\text{e}}},
\end{equation}
and
\begin{equation}
\label{eq:udampP}
\omega_\text{P} = \frac{1}{4c}\left(-iA_\text{e/m} \pm \sqrt{16c^2\omega_\text{e/m}^2-A_\text{e/m}^2}\right).
\end{equation}
In this situation, the complex frequencies of the transmission Zeros read
\begin{equation}
\label{eq:udampZ}
    \omega_\text{Z} = \pm \frac{1}{2\sqrt{2}c}\sqrt{4c^2\left(\omega_\text{e}^2+\omega_\text{m}^2\right)-A_\text{e}A_\text{m} \pm B},
\end{equation}
where
\begin{equation}
\label{eq:Bparam}
    B = \sqrt{[4c^2\left(\omega_\text{e}+\omega_\text{m}\right)^2-A_\text{e}A_\text{m}][4c^2\left(\omega_\text{e}-\omega_\text{m}\right)^2-A_\text{e}A_\text{m}]}.
\end{equation}

We now focus on the trajectories of transmission zeros for the case of undamped metasurface discussed previously in Fig. \ref{fig:EP}. For this purpose, we use~\eqref{eq:udampZ} to predict the branching behavior already observed above where transmission Zeros bifurcate from real to complex values. This happens when the term under the square root in the expression of B is negative, satisfied at the condition:
\begin{equation}
\label{eq:ROIcond}
    (\omega_\text{e} - \omega_\text{m})^2 < \frac{A_\text{e}A_\text{m}}{4c^2}.
\end{equation}
The exceptional points then occur at the boundary of this region where B becomes null:
\begin{equation}
\label{eq:Zmergecond}
    (\omega_\text{e} - \omega_\text{m})^2 = \frac{A_\text{e}A_\text{m}}{4c^2}.
\end{equation}
The exceptional point positions are then provided by the following expression:
$\omega = \pm\sqrt{\omega_\text{e}\omega_\text{m}}$.
Looking at the asymptotic behaviors of Poles and transmission Zeros, we can note in Fig. \ref{fig:phase}(c) that the real frequencies of the Poles and Zeros tend toward each other far from the region of interest, i.e., $h > 250$~nm and $h< 100$~nm or when condition~\eqref{eq:ROIcond} is not satisfied. This is verified by neglecting the contributions from $A_\text{e}A_\text{m}$ in~\eqref{eq:udampP} and~\eqref{eq:udampZ}, which yields
\begin{subequations}
    \label{eq:PZasymp}
    \begin{align}
        \omega_\text{P} &\approx\pm \omega_\text{e/m}- \frac{i}{4c}A_\text{e/m},\\
        \omega_\text{Z} &\approx\pm \omega_\text{e/m}.
    \end{align}
\end{subequations}


\end{document}